\documentclass[reprint,prc,showpacs,amsmath,amssymb,superscriptaddress]{revtex4-1}
\usepackage{dcolumn}
\usepackage[dvips]{graphicx}
\usepackage{amsmath}
\usepackage{multirow}
\usepackage{setspace}
\newcommand\T{\rule{0pt}{3.1ex}}
\usepackage{color}
\usepackage{bm}
\usepackage[T1]{fontenc}
\usepackage[latin9]{inputenc}
\usepackage{color}
\usepackage{amssymb}
\usepackage{float}

\providecommand{\tabularnewline}{\\}

\usepackage{array}

\begin{document}

\title{Neutrinoless double positron decay and positron emitting electron capture in the interacting boson model}

\author{J.\ Barea}
\email{jbarea@udec.cl}
\affiliation{Departamento de F\'{i}sica, Universidad de Concepci\'{o}n,
 Casilla 160-C, Concepci\'{o}n 4070386, Chile}

\author{J. Kotila}
\email{jenni.kotila@yale.edu}
\affiliation{Center for Theoretical Physics, Sloane Physics Laboratory,
 Yale University, New Haven, Connecticut 06520-8120, USA}

\author{F.\ Iachello}
\email{francesco.iachello@yale.edu}
\affiliation{Center for Theoretical Physics, Sloane Physics Laboratory,
 Yale University, New Haven, Connecticut 06520-8120, USA}

\begin{abstract}
Neutrinoless double-$\beta$ decay is of fundamental importance for determining the neutrino mass. Although double electron ($\beta^-\beta^-$) decay is the most promising mode, in very recent years interest in double positron ($\beta^+\beta^+$) decay, positron emitting electron capture ($EC\beta^+$), and double electron capture ($ECEC$) has been renewed. We present here results of a calculation of nuclear matrix elements for neutrinoless double-$\beta^+$ decay and positron emitting electron capture within the framework of the microscopic interacting boson model (IBM-2) for $^{58}$Ni, $^{64}$Zn, $^{78}$Kr, $^{96}$Ru, $^{106}$Cd, $^{124}$Xe, $^{130}$Ba, and $^{136}$Ce decay. By combining these with a calculation of phase space factors we calculate expected half-lives.  

\end{abstract}


\maketitle

\section{Introduction}

Double-$\beta$ decay is a process in which a nucleus $(A,Z)$ decays to a nucleus $(A,Z\pm2)$ by emitting two electrons or positrons and, usually, other light particles
\begin{equation}
(A,Z)\rightarrow(A,Z\pm 2) + 2e^{\mp} + \text{anything}.
\end{equation}
Double-$\beta$ decay can be classified in various modes according to the various types of particles emitted in the decay. The processes where two neutrinos are emitted are predicted by the standard model, and $2\nu\beta^-\beta^-$ decay has been observed in several nuclei.
 For processes not allowed by the standard model, i.e. the neutrinoless modes: $0\nu\beta\beta$, $0\nu\beta EC$, $0\nu ECEC$, the half-life can be factorized as 
\begin{equation}
\label{0nut}
\left[\tau^{0\nu}_{1/2}\right]^{-1}=G_{0\nu}|M_{0\nu}|^2 \left| f(m_i,U_{ei})\right|^2,
\end{equation}
where $G_{0\nu}$ is a phase space factor, $M_{0\nu}$ is the nuclear matrix element, and $f(m_i, U_{ei})$ contains physics beyond the standard model through the masses $m_i$ and mixing matrix elements $U_{ei}$ of neutrino species. For all processes, two crucial ingredients are the phase space factors (PSFs) and the nuclear matrix elements (NMEs). Recently, we have initiated a program for the evaluation of both quantities and presented results for $\beta^-\beta^-$ decay \cite{barea09,kotila12,barea12,barea12b, kotila12b, barea12c, fink12}. 
This is the most promising mode for the possible detection of neutrinoless double-$\beta$ decay and thus of a measurement of the absolute neutrino mass scale. However, in very recent years, interest in the double positron decay, $\beta^+\beta^+$, positron emitting electron capture, $EC\beta^+$, and double electron capture $ECEC$, has been renewed. This is due to the fact that positron emitting processes have interesting signatures that could be detected experimentally \cite{zuber}. In a previous article \cite{kot12b} we initiated a systematic study of  $\beta^+\beta^+$,  $EC\beta^+$, and  $ECEC$ processes and  presented a calculation of phase space factors (PSF) for $2\nu\beta^+\beta^+$, $2\nu EC\beta^+$, $2\nu ECEC$ and $0\nu\beta^+\beta^+$, $2\nu EC\beta^+$. The process $0\nu ECEC$ cannot occur to the order of approximation used in \cite{kot12b}, since the emission of additional particles, $\gamma\gamma$ or others, is needed to conserve energy and momentum. In this article, we focus on calculation of neutrinoless decay nuclear matrix elements (NME), which are common to all three modes, and half-life predictions for $0\nu\beta^+\beta^+$ and $0\nu EC\beta^+$ modes. Results of our calculations are reported for nuclei listed in Table~\ref{table1}.

\begin{ruledtabular}
\begin{center}
\begin{table*}[cbt!]
\caption{\label{table1}Double-$\beta$ decays considered in this article, the mass difference between neutral mother and daughter atoms, $M(A,Z)-M(A,Z-2)$, and their isotopic abundances.}
\begin{tabular}{lll}
transition  &$M(A,Z)-M(A,Z-2)$(keV)\footnotemark[1]&$P$(\%) \\
\hline
\T
$_{28}^{58}$Ni$_{30}\rightarrow _{26}^{58}$Fe$_{32}$ &$1926.3 \pm 0.3$  	&$68.077\pm 0.009$ \\
$_{30}^{64}$Zn$_{34}\rightarrow _{28}^{64}$Ni$_{36}$ &$1094.8 \pm 0.7$  	&$49.17\pm 0.75$\\
$_{36}^{78}$Kr$_{42}\rightarrow _{34}^{78}$Se$_{44}$ &$2846.3 \pm 0.7$  	&$0.355\pm 0.003$ \\
$_{44}^{96}$Ru$_{52}\rightarrow _{42}^{96}$Mo$_{54}$ &$2714.51 \pm 0.13$\footnotemark[2]  	&$5.54\pm 0.14$ \\
$_{48}^{106}$Cd$_{58}\rightarrow _{46}^{106}$Pd$_{60}$ &$2775.39 \pm 0.10$\footnotemark[3]  &$1.25\pm 0.06$ \\
$_{54}^{124}$Xe$_{70}\rightarrow _{52}^{124}$Te$_{72}$ &$2865.4 \pm 2.2$  	&$0.0952\pm 0.0003$ \\
$_{56}^{130}$Ba$_{74}\rightarrow _{54}^{130}$Xe$_{76}$ &$2619 \pm 3$  		&$0.106\pm 0.001$ \\
$_{58}^{136}$Ce$_{78}\rightarrow _{56}^{136}$Ba$_{80}$ &$2378.53 \pm 0.27$\footnotemark[4]  &$0.185\pm 0.002$ \\
\end{tabular}
\footnotetext[1]{Ref.~\cite{nudat}}
\footnotetext[2]{Ref.~\cite{prc83}}
\footnotetext[3]{Ref.~\cite{prc84}}
\footnotetext[4]{Ref.~\cite{plb697}}
\end{table*}
\end{center}
\end{ruledtabular}

\section{Results}
\subsection{Nuclear matrix elements}

The theory of 0$\nu\beta\beta$ decay was first formulated by Furry
\cite{furry} and further developed by Primakoff and Rosen \cite{primakoff},
Molina and Pascual \cite{molina}, Doi \textit{et al.} \cite{doi},
Haxton and Stephenson \cite{haxton}, and, more recently, by Tomoda
\cite{tomoda} and \v{S}imkovic \textit{et al.} \cite{simkovic}. All
these formulations often differ by factors of 2, by the number of
terms retained in the non-relativistic expansion of the current and
by their contribution. In order to have a standard set of calculations
to be compared with the QRPA and the ISM, we adopt in this article the formulation
of \v{S}imkovic \textit{et al.} \cite{simkovic}. A detailed discussion of involved operators can also be found in Ref. \cite{barea12b}.

\begin{figure}[h]
\begin{center}
\includegraphics[width=8.6cm]{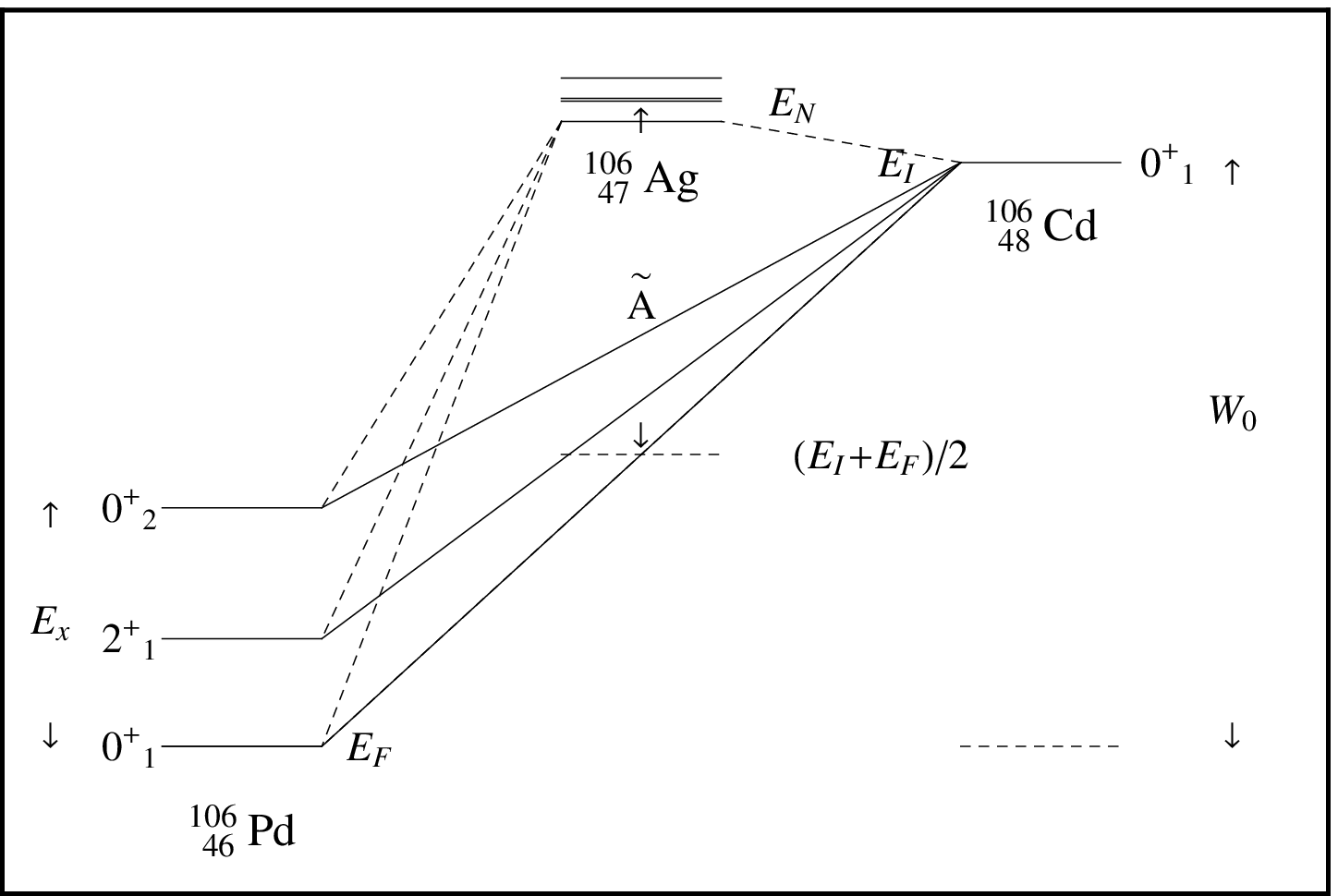} 
\end{center}
\caption{\label{fig1} The decay $_{106}^{48}$Cd$_{58}  \rightarrow _{106}^{46}$Pd$_{60}$, an example of double-$\beta^+$ decay.}
\end{figure}

We consider the decay of a nucleus $_{Z}^{A}$X$_{N}$ into a nucleus
$_{Z-2}^{A}$Y$_{N+2}$. An example is shown in Fig.~\ref{fig1}. 
If the decay proceeds through an $s$-wave, with two leptons in the
final state, we cannot form an angular momentum greater than one. We
therefore calculate, in this article, only $0\nu\beta\beta$ matrix elements to final
$0^{+}$ states, the ground state $0_{1}^{+}$, for which, in a previous article \cite{kot12b}
we have calculated the phase space factors, and to the first excited
state $0_{2}^{+}$.

In order to evaluate the matrix elements we make use of the microscopic
interacting boson model (IBM-2) \cite{iac1}. The method of evaluation
is discussed in detail in Ref.~\cite{barea09} for double electron decay ($\beta^-\beta^-$). For double positron decay ($\beta^+\beta^+$) and positron emitting electron capture ($EC\beta^+$) the same method applies except for the interchange $\pi \rightarrow \nu$ in Eq.~(5) of \cite{barea09} and in the mapped boson operators of Eq.~(18) of \cite{barea09}. The matrix elements of the mapped operators are evaluated with
realistic wave functions, taken either from the literature, when available,
or obtained from a fit to the observed energies and other properties
($B(E2)$ values, quadrupole moments, $B(M1)$ values, magnetic moments, etc.).
The values of the parameters used in the calculation are given in Appendix A.

\begin{ruledtabular}
\begin{table*}[cbt!]
\caption{\label{table2}IBM-2 nuclear matrix elements $M^{(0\nu)}$ (dimensionless) for neutrinoless $\beta^{+}\beta^{+}$/$EC\beta^{+}$/$ECEC$ decay with Jastrow M-S SRC and $g_V/g_A=1/1.269$.}
\begin{tabular}{lcccccccc}
Nucleus & \multicolumn{4}{c}{$0_{1}^{+}$} & \multicolumn{4}{c}{$0_{2}^{+}$}\tabularnewline \cline{2-5} \cline{6-9}
\T
& $M_{GT}^{(0\nu)}$ & $M_{F}^{(0\nu)}$  & $M_{T}^{(0\nu)}$ & $M^{(0\nu)}$ & $M_{GT}^{(0\nu)}$ & $M_{F}^{(0\nu)}$  & $M_{T}^{(0\nu)}$ & $M^{(0\nu)}$\tabularnewline
\hline 
\T
$^{58}$Ni  & 2.072 & -0.152 & 0.144  & 2.310 		& 2.042 & -0.153 & 0.101 &  2.237\tabularnewline
$^{64}$Zn  & 4.762 & -2.449 & -0.156 & 6.127 		& 0.633 & -0.360 & -0.019 & 0.837\tabularnewline
$^{78}$Kr  & 3.384 & -2.146 & -0.238 & 4.478 		& 0.771 & -0.479 & -0.055 & 1.014\tabularnewline
$^{96}$Ru  & 2.204 & -0.269 & 0.112  & 2.483 		& 0.036 & -0.012 & 0.001 & 0.045\tabularnewline
$^{106}$Cd & 2.757 & -0.255 & 0.191  & 3.106 		& 1.395 & -0.110 & 0.074 & 1.537\tabularnewline
$^{124}$Xe & 3.967 & -2.224 & -0.192 & 5.156 		& 0.647 & -0.359 & -0.032 & 0.839\tabularnewline
$^{130}$Ba & 3.911 & -2.108 & -0.176 & 5.043 		& 0.285 & -0.152 & -0.014 & 0.366\tabularnewline
$^{136}$Ce & 3.815 & -2.007 & -0.161 & 4.901 		& 0.318 & -0.167 & -0.014 & 0.408\tabularnewline
\end{tabular}
\end{table*}
\end{ruledtabular}

Here, we present our calculated NME for the decays of Table~\ref{table1}. The NMEs depend on many assumptions, in particular on the treatment  of the short-range correlations (SRC). In Table~\ref{table2}, we show the results of our calculation of the matrix
elements to the ground state, $0_{1}^{+}$, and to the first excited state,
$0_{2}^{+}$, using the Miller-Spencer (MS) parametrization of SRC, and broken down into GT, F and T contributions and their sum as
\begin{equation}
\begin{split}
M_{0\nu} & =  g_{A}^{2}M^{(0\nu)}, \\
M^{(0\nu)} & =  M_{GT}^{(0\nu)} -\left(\frac{g_{V}}{g_{A}}\right)^2 M_{F}^{(0\nu)}+M_{T}^{(0\nu)}.
\end{split}
\end{equation}
We note that we have two classes of nuclei, those in which protons and neutrons occupy the same major shell ($A=64,78, 124, 128, 130, 136$) and those in which they occupy different major shells ($A=58, 96, 106$). The magnitude of the Fermi matrix element, which is related to the overlap of the proton and neutron wave functions, is therefore different in these two classes of nuclei, being large in the former and small in the latter case. This implies a considerable amount of isospin violation for nuclei in the first class. This problem has been discussed in detail in Ref.~\cite{barea12b} and will form a subject of subsequent investigation. It is common to most calculations of NME and has been addressed recently within the framework of QRPA in Refs.~\cite{rod11,sim13}. Here we take it into account by assigning a large error to the calculation of the Fermi matrix elements. In the same Ref.~\cite{barea12b} it is also shown that the NME depend on the short range correlations (SRC), and that use of Argonne/CD-Bonn SRC increases the NME by a factor of 1.1-1.2. The same situation occurs for $\beta^+\beta^+$ decay. In order to take into account the sensitivity of the calculation to parameter changes, model assumptions and operator assumptions \cite{barea12b}, we list in Table~\ref{final} IBM-2 NMEs with an estimate of the error. 
The values of the $0_1^+$ matrix elements vary between $~2.3-6.1$, the matrix element for the $^{64}$Zn$\to ^{64}$Ni transition being notably the largest. They are therefore of the same order of magnitude than the nuclear matrix elements for $\beta^-\beta^-$ decay, 2.0-5.4. 

In the same Table~\ref{final} we also compare our results with the available QRPA calculations from Ref. \cite{hir94} with the addition of some more recent calculations from Refs.~\cite{suh12, suh11}. The QRPA~\cite{hir94} NMEs are calculated taking into account GT and F contributions, and using the value $g_A=1.25$.
As in the case of $\beta^-\beta^-$ decay, QRPA tend to give larger values than IBM-2 and these two methods seem to be in a rather good correspondence with each other.

\begin{ruledtabular}
\begin{table}[h]
\caption{\label{final}IBM-2 matrix elements with M-S SRC and error estimate compared with available QRPA calculations.}
\begin{tabular}{lccccc}
Decay &\multicolumn{3}{c}{$0_{1}^{+}$} &\multicolumn{2}{c}{$0_{2}^{+}$}\tabularnewline \cline{2-4} \cline{5-6}
\T
&IBM-2 &\multicolumn{2}{c}{QRPA\footnotemark[1]} &IBM-2 &QRPA\\ 
\hline 
\T
$^{58}$Ni  & 2.31(37) 	&1.55	&	& 2.24(36) &\tabularnewline
$^{64}$Zn  & 6.13(116) 	&	&	& 0.84(16) &\tabularnewline
$^{78}$Kr  & 4.48(85) 	&4.19	&	& 1.01(19) &\tabularnewline
$^{96}$Ru  & 2.48(40) 	&3.25	&3.22-5.83\footnotemark[2]	& 0.05(1) &1.28-2.26\footnotemark[2]\tabularnewline
$^{106}$Cd & 3.11(50) 	&4.12	&5.94-9.08\footnotemark[3]	& 1.54(25) &0.66-0.91\footnotemark[3]\tabularnewline
$^{124}$Xe & 5.16(98) 	&4.78	&	& 0.84(16) &\tabularnewline
$^{130}$Ba & 5.04(96) 	&4.98	&	& 0.37(7) &\tabularnewline
$^{136}$Ce & 4.90(93) 	&3.09	&	& 0.41(8) &\tabularnewline
\end{tabular}
\footnotetext[1]{Reference~\cite{hir94}.}
\footnotetext[2]{Reference~\cite{suh12}.}
\footnotetext[3]{Reference~\cite{suh11}.}
\end{table}
\end{ruledtabular}

\subsection{Predicted half-lives for $0^+_{1}\to 0^+_{1}$ transitions}
\begin{figure}[h]
\begin{center}
\includegraphics[width=8.6cm]{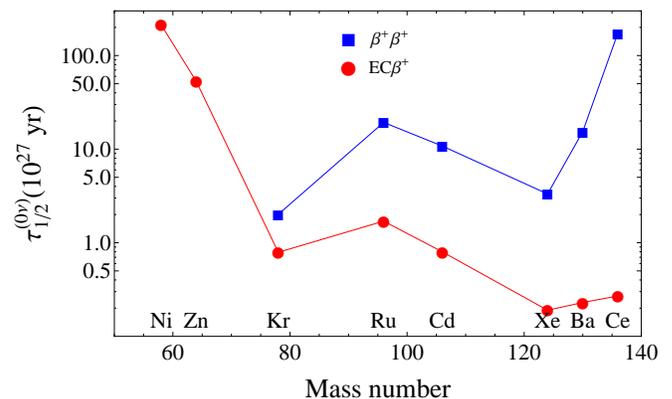} 
\end{center}
\caption{\label{fig2}(Color online) Expected half-lives for $\left\langle m_{\nu}\right\rangle=1$~eV, $g_{A}=1.269$.  The figure is in semilogarithmic scale.}
\end{figure}

The calculation of nuclear matrix elements in IBM-2 can now be combined
with the phase space factors calculated in \cite{kot12b} to produce our final results
for half-lives for light neutrino exchange in Table~\ref{table3} and Fig.~\ref{fig2}. The half-lives are calculated using the formula 
\begin{equation}
\lbrack\tau_{1/2}^{0\nu}]^{-1}=G_{0\nu}^{i}\left\vert M_{0\nu}\right\vert ^{2}\left\vert \frac{\left\langle m_{\nu}\right\rangle }{m_{e}}\right\vert ^{2},
\end{equation}
where $i=\beta^+\beta^+, EC\beta^+$.
The values in Table~\ref{table3} and Fig.~\ref{fig2} are for $\left\langle m_{\nu}\right\rangle = 1$eV. They scale with $\left\langle m_{\nu}\right\rangle^2$ for other values.

\begin{ruledtabular}
\begin{table}[h]
\caption{\label{table3}Calculated half-lives in IBM-2 M-S SRC for neutrinoless double-$\beta^+$ decay and positron emitting electron capture for $\left<m_{\nu}\right>=1$~eV and $g_A=1.269$.}
\begin{tabular}{lcc}
&\multicolumn{2}{c}{$T_{1/2}$($10^{27}$yr)}\\
Nucleus &$\beta^+\beta^+$ &$EC\beta^+$ \\
\hline
\T
$^{58}$Ni	&			&213	\\
$^{64}$Zn	&			&52.9	\\
$^{78}$Kr	&2.01		&0.79	\\
$^{96}$Ru	&19.3		&1.70	\\
$^{106}$Cd	&10.8		&0.80	\\
$^{124}$Xe	&3.32		&0.19	\\
$^{130}$Ba	&15.4		&0.23	\\
$^{136}$Ce	&174		&0.27	\\
\end{tabular}
\end{table}
\end{ruledtabular}

Comparing the half-life predictions listed in Table~\ref{table3} to the ones reported in Ref.~\cite{barea12b} for $0\nu\beta^-\beta^-$ we can see that values reported here are much larger. This is due to the fact that in cases studied here the available kinetic energy is much smaller compared to $\beta^-\beta^-$ decay. Furthermore,  the Coulomb repulsion on positrons from the nucleus gives a smaller decay rate. As concluded also in Refs.~\cite{doi93, kim83}, the $^{124}$Xe $0\nu EC\beta^+$decay is expected to have the shortest half-live. 
In case of the neutrinoless double electron capture process, $0\nu ECEC$, the available kinetic energy is larger and Coulomb repulsion does not play a role. However, this decay mode cannot occur to the order of approximation we are considering, since it must be accompanied by the emission of one or two particles in order to conserve energy, momentum and angular momentum.

\section{Conclusions}
In this article we have presented evaluation of nuclear
matrix elements in 0$\nu\beta^+\beta^+$/0$\nu EC\beta^+$/0$\nu ECEC$ within the framework of IBM-2 in the closure approximation. The closure approximation  is expected to be good for these decays since the virtual neutrino momentum is of order 100 MeV/c and thus much larger than the scale of nuclear excitations. 
By using these matrix elements and the phase space factors of Ref.~\cite{kot12b}, we have calculated the expected 0$\nu\beta^+\beta^+$/0$\nu EC\beta^+$ half-lives in all nuclei of interest with $g_A=1.269$ and $g_V=1$, given in Table~\ref{table3} and Fig.~\ref{fig2}.

\begin{acknowledgments}
This work was performed in part under the US DOE Grant DE-FG-02-91ER-40608 and Fondecyt Grant No. 1120462. We wish to thank 
K. Zuber for stimulating discussions.
\end{acknowledgments}

\section{Appendix A}

A detailed description of the IBM-2 Hamiltonian is given in \cite{iac1} and \cite{otsukacode}. For most nuclei, the Hamiltonian parameters are taken from the literature \cite{zn, Kaup79, Kaup83, Shlomo92, Sambataro82,kim96, Puddu80}. The values of the Hamiltonian parameters, as well as the references from which they were taken, are given in Table~\ref{tab:ibm2parameters}. The quality of the description can be seen from these references and ranges from very good to excellent. 
\begin{table*}[ctb!]
 \caption{Hamiltonian parameters employed in the IBM-2 calculation of the final
wave functions along with their references.}
\label{tab:ibm2parameters}
\begin{ruledtabular}
\begin{centering}
\begin{tabular}{lcccccccccccccc}
Nucleus  & $\epsilon_{d_{\nu}}$  & $\epsilon_{d_{\pi}}$  & $\kappa$  & $\chi_{\nu}$  & $\chi_{\pi}$  & $\xi_{1}$  & $\xi_{2}$  & $\xi_{3}$  & $c_{\nu}^{(0)}$  & $c_{\nu}^{(2)}$  & $c_{\nu}^{(4)}$  & $c_{\pi}^{(0)}$  & $c_{\pi}^{(2)}$  & $c_{\pi}^{(4)}$         \tabularnewline
\hline 
\T
$^{58}\mbox{Ni}$\footnotemark[1]  & 1.454  &   &  &  &  &   &   &   &  &  &  &  &  &   \tabularnewline
$^{58}\mbox{Fe}$ \footnotemark[1]  & 0.98  & 0.98  & -0.26  & 0.00  & -0.40  & 0.80  & 0.80  & 0.80  &  &  &  &  &  &   \tabularnewline
$^{64}\mbox{Zn}$ \cite{zn}  &1.20   &1.20   &-0.22   &-0.25   &-0.75   &-0.18  &0.24  &-0.18   &-0.30  &-0.50  &0.30  &-0.30  &-0.50  &0.30   \tabularnewline
$^{64}\mbox{Ni}$ \footnotemark[1]  & 1.346  &  &  &  &   &  &  &   &  &-0.415  &0.082  &  &  &  \tabularnewline
$^{78}\mbox{Kr}$ \cite{Kaup79}  & 0.96  & 0.96  & -0.18  & -0.495  & -1.127  &-0.10   &  &-0.10   &  &  &  &  &  &\tabularnewline
$^{78}\mbox{Se}$ \cite{Kaup83}  & 0.99  & 0.99  & -0.21  & 0.71  & -0.90  &  &  &-0.10  &  &  &  &  &  &   \tabularnewline
$^{96}\mbox{Ru}$ \footnotemark[1]  & 1.08  & 1.08  & -0.21  & 0.80  & 0.40  &0.25  & 0.25  & 0.25  & 0.30  & 0.10  &-0.50  &  &  &   \tabularnewline
$^{96}\mbox{Mo}$ \cite{Shlomo92}  & 0.73  & 1.10  & -0.09  & -1.20  & 0.40  & -0.10  & 0.10  & -0.10  & -0.50  & 0.10  &  &  &  &   \tabularnewline
$^{106}\mbox{Cd}$ \cite{Sambataro82}  & 1.05  & 1.05  & -0.325  & 1.25  &0.00  &-0.18  &0.24  &-0.18  & 0.20  & 0.15  &0.00  &  &  &   \tabularnewline
$^{106}\mbox{Pd}$ \cite{kim96}  & 0.760  & 0.844  & -0.160  & -0.22  & -0.30  &0.20  &0.05  &0.00  & -0.45  & -0.20  & 0.01  &   &   &    \tabularnewline


$^{124}\mbox{Xe}$ \cite{Puddu80}  & 0.70  & 0.70  & -0.14  & 0.00  & -0.80  & -0.18  & 0.24  &-0.18  & 0.05  & -0.16  &  &  &  &   \tabularnewline
$^{124}\mbox{Te}$ \cite{Sambataro82}  & 0.82  & 0.82  & -0.15  & 0.00  & -1.20  & -0.18  & 0.24  & -0.18  &0.10  &  &  &  &  &   \tabularnewline
$^{130}\mbox{Ba}$ \cite{Puddu80}  & 0.70  &0.70  &-0.175  &0.32  &-0.90  &-0.18  &0.24  &-0.18  & 0.26  &  &  &  &  &  \tabularnewline
$^{130}\mbox{Xe}$ \cite{Puddu80}  & 0.76  &0.76  &-0.19  &0.50  &-0.80  &-0.18  &0.24  &-0.18  & 0.30  & 0.22  &   &  &  &  \tabularnewline
$^{136}\mbox{Ce}$ \cite{Puddu80}  & 0.90  & 0.90  & -0.21  & 0.79  & -1.00  & -0.18  & 0.24  & -0.18  & 0.26  &-0.11  &  &  &  &   \tabularnewline
$^{136}\mbox{Ba}$ \cite{Puddu80}  & 1.03  & 1.03  & -0.23  & 1.00  & -0.90  & -0.18  & 0.24  & -0.18  & 0.30  & 0.10  &  &  &  &   \tabularnewline
\end{tabular}
\footnotetext[1]{Parameters fitted to reproduce the spectroscopic
data of the low lying energy states.}
\par\end{centering}
 \end{ruledtabular} 
\end{table*}


\end{document}